# CRi
## Computer Law Review International
A Journal of Information Law and Technology

## Articles

*Michael Veale / Frederik Zuiderveen Borgesius\**

# Demystifying the Draft EU Artificial Intelligence Act
Analysing the good, the bad, and the unclear elements of the proposed approach

*In April 2021, the European Commission proposed a Regulation on Artificial Intelligence, known as the AI Act. We present an overview of the Act and analyse its implications, drawing on scholarship ranging from the study of contemporary AI practices to the structure of EU product safety regimes over the last four decades. Aspects of the AI Act, such as different rules for different risk-levels of AI, make sense. But we also find that some provisions of the Draft AI Act have surprising legal implications, whilst others may be largely ineffective at achieving their stated goals. Several overarching aspects, including the enforcement regime and the risks of maximum harmonisation pre-empting legitimate national AI policy, engender significant concern. These issues should be addressed as a priority in the legislative process.*

## I. Introduction

1 On 21 April 2021, the European Commission presented a proposal for a Regulation concerning artificial intelligence (AI), – the AI Act, for short.[1] This Draft AI Act seeks to lay down harmonised rules for the development, placement on the market and use of AI systems which vary by characteristic and risk, including prohibitions and a conformity assessment system adapted from EU product safety law.

2 In this paper, we analyse the initial Commission proposal – the first stage in a potentially long law-making process.[2] The Draft AI Act is sufficiently complex to prevent us from summarising it exhaustively. We instead aim to contextualise and critique it, and increase accessibility of the debate to stakeholders who may struggle to apply their expertise and experience to what at times can be an arcane proposal.

## 1. Context

3 The first public indication of regulatory action of the type proposed in the Draft AI Act were a cryptic few sentences found in the previous European Commission's contribution to the Sibiu EU27 leader's meeting in 2019.[3] Subsequently, then-President-Elect von der Leyen's political guidelines for the Commission indicated an intention to 'put forward legislation for a coordinated European approach on the human and ethical impli-cations of Artificial Intelligence'⁴ – the spark that the Draft AI Act acknowledges as its genesis.⁵ The proposed Regulation is part of a tranche of proposals which must be understood in tandem, including:

- the draft Digital Services Act (with provisions on recommenders and research data access);⁶
- the draft Digital Markets Act (with provisions on AI-relevant hardware, operating systems and software distribution);⁷
- the draft Machinery Regulation⁸ (revising the Machinery Directive in relation to AI, health and safety, and machinery);
- announced product liability revision relating to AI;⁹
- the draft Data Governance Act (concerning data sharing frameworks).¹⁰

---

\* Thanks to Valerio De Stefano, Reuben Binns, Jeremias Adams-Prassl, Barend van Leeuwen, Aislinn Kelly-Lyth, Lilian Edwards, Natali Helberger, Christopher Marsden, Sarah Chander, Corinne Cath-Speth for comments and/or discussion; substantive and editorial input by Ulrich Gasper; and the conveners and participants of several workshops including one convened by Margot Kaminski, one by Burkhard Schäfer, one part of the 2nd ELLIS Workshop in Human-Centric Machine Learning; one between Lund University and the Labour Law Community; and one between Oxford, KU Leuven and UCL. A CC-BY 4.0 license applies to this article after 3 calendar months from publication have elapsed.

1 European Commission, Proposal for a Regulation of the European Parliament and of the Council laying down harmonised rules on artificial intelligence (Artificial Intelligence Act) and amending certain Union legislative acts (COM(2021) 206 final) (hereafter 'Draft AI Act').

2 For readers unfamiliar with the European legislative process: the Commission is the European Union's executive, and has a monopoly on policy initiative. Drafts are amended and adopted through a bicameral procedure between the directly elected European Parliament and the Council, which represents Member State governments. This procedure encompasses both formal stages and informal back-room compromise (trialogue).

3 European Commission, 'Europe in May 2019: Preparing for a More United, Stronger and More Democratic Union in an Increasingly Uncertain World' (Contribution to the informal EU27 leaders' meeting in Sibiu (Romania) on 9 May 2019, 9 May 2019) 33.

## 2. Structure and Approach

4 The 'Act' is a regulation based on Article 114 of the Treaty on the Functioning of the European Union (TFEU), which concerns the approximation of laws to improve the functioning of the internal market. The proposal mixes reduction of trade barriers with broad fundamental rights concerns in a structure unfamiliar to many information lawyers, and with significant consequences on the space for Member State action which we discuss further below. While it may look new, much of the Draft AI Act's wording is drawn from a 2008 Decision establishing a framework for certain regulations concerning product safety, used in a wide array of subsequent legislation.¹¹ The main enforcement bodies of the proposed AI Act, 'market surveillance authorities' (MSAs), are also common in EU product law. All this brings a range of novelties and tensions we will explore.

5 The Commission distinguishes different risk levels regarding AI practices, which we adapt to analyse in four categories: i) *unacceptable risks* (Title II); ii) *high risks* (Title III); iii) *limited risks* (Title IV); iv) *minimal risks* (Title IX). We cover each in turn, except for *minimal risks*, where Member States and the Commission merely 'encourage' and 'facilitate' voluntary codes of conduct.¹² We finally look at broader themes raised by the Draft AI Act, in particular the important question of pre-emption and residual competences of Member States, and enforcement.

## II. Title II: Unacceptable Risks

6 Unacceptable risks attract outright or qualified prohibitions in the Draft AI Act. Whether the AI Act would contain prohibited practices has been a matter of controversy. In 2018, the Commission set up a 'High-Level Expert Group on AI' to advise on its AI strategy. Members soon described industry pressure that led to the group dropping terms including 'red lines' and 'non-negotiable' from their policy recommendations.¹³ A leaked version of the Commission *White Paper on Artificial Intelligence* contained a moratorium on facial recognition, controversially later expunged from the final version.¹⁴

7 The Commission's proposal contains four prohibited categories, three prohibited in their entirety (two on manipulation, one on social scoring); and the last, 'real-time' and 'remote' biometric identification systems prohibited except for specific law enforcement purposes if accompanied by an independent authorisation regime.

### 1. Manipulative Systems

8 Two prohibited practices claim to regulate manipulation.¹⁵

**(a)** the placing on the market, putting into service or use of an AI system that deploys <u>subliminal techniques</u> beyond a person's consciousness <u>in order to</u> materially distort a person's behaviour in a manner that causes or is likely to cause that person or another person <u>physical or psychological harm</u>;

**(b)** the placing on the market, putting into service or use of an AI system that exploits any of the vulnerabilities of a specific group of persons due to their <u>age</u>, <u>physical or mental disability</u>, <u>in order to</u> materially distort the behaviour of a person pertaining to that group in a manner that causes or is likely to cause that person or another person <u>physical or psychological harm</u>; (emphases added)

9 In briefings on the prohibitions, the Commission has presented an example for each. They border on the fantastical. A crossover episode of *Black Mirror* and the Working Time Directive exemplifies the first: '[a]n inaudible sound [played] in truck drivers' cabins to push them to drive longer than healthy and safe [where] AI is used to find the frequency maximising this effect on drivers'. The second is a '[a] doll with integrated voice assistant [which] encourages a minor to engage in progressively

---

4 Ursula von der Leyen, 'A Union that Strives for More: My Agenda for Europe' (Political Guidelines for the Next European Commission 2019-2024, 2019).

5 AI Act, p.1.

6 European Commission, Proposal for a Regulation of the European Parliament and of the Council on a Single Market For Digital Services (Digital Services Act) and amending Directive 2000/31/EC (COM(2020) 825 final).

7 European Commission, Proposal for a Regulation of the European Parliament and of the Council on contestable and fair markets in the digital sector (Digital Markets Act) (COM(2020) 842 final).

8 European Commission, Proposal for a Regulation of the European Parliament and of the Council on machinery products (COM(2021) 202 final) (Machinery Regulation).

9 See European Commission, 'Communication from the Commission to the European Parliament, the Council, the European Economic and Social Committee and the Committee of the Regions: Fostering a European Approach to Artificial Intelligence (COM(3032) 205 Final)' (21 April 2021) 2.

10 European Commission, Proposal for a Regulation of the European Parliament and of the Council on European Data Governance (Data Governance Act) (COM(2020) 767 final).

11 Decision No 768/2008/EC of the European Parliament and of the Council of 9 July 2008 on a common framework for the marketing of products, and repealing Council Decision 93/465/EEC, OJ L 218/82.

12 AI Act, art 69.

13 Thomas Metzinger, 'Ethics Washing Made in Europe', *Der Tagesspiegel* (18 April 2019) https://www.tagesspiegel.de/politik/eu-guidelines-ethics-washing-made-in-europe/24195496.html accessed 15 August 2019. Two-thirds of HLEG-AI members were industry representatives. See generally Michael Veale, 'A Critical Take on the Policy Recommendations of the EU High-Level Expert Group on Artificial Intelligence' [2020] European Journal of Risk Regulation.

14 Access Now, 'Europe's Approach to Artificial Intelligence: How AI Strategy is Evolving' (December 2020) 24–25 https://perma.cc/X3JM-2M6A.

15 AI Act, arts 5(1)(a–b); described as manipulation in recital 15; pages 12–13.

dangerous behavior or challenges in the guise of a fun or cool game'.[16]

10  These provisions jar with a common understanding of manipulation. Manipulation can be understood through four necessary, cumulative conditions: the manipulator wants to *intentionally* but *covertly* make use of another's decision-making to *further their own ends* through exploiting some *vulnerability* (understood broadly).[17] The Draft AI Act's provisions echo some of these conditions. The Draft AI Act requires *intent* ('in order to'). It is limited to certain *vulnerabilities*, either caused by 'age, physical and mental disability' or exposed through 'subliminal techniques'. If reliant on subliminal techniques, they must be *covert* ('beyond a person's consciousness'). However, a final trigger is not whether a would-be manipulator's own ends are furthered, but instead on whether the activity 'causes or is likely to cause that person or another person physical or psychological harm'. This heavily limits the provision's scope.

### a) The Harm Requirement

11  Manipulative AI systems appear permitted insofar as they are unlikely to cause an individual (not a collective) 'harm'. This harm requirement entails a range of problematic loopholes. A cynic might feel the Commission is more interested in prohibitions' rhetorical value than practical effect.

12  In real life, harm can accumulate without a single event tripping a threshold of seriousness, leaving it difficult to prove.[18] These 'cumulative' harms are reinforced over time by their impact on individuals' environments, with hyperpersonalisation, engagement and 'dwell' metrics and impact on children often called out in this regard.[19] Indeed, manipulation in other fields of law leaves the Draft AI Act already looking dated. Law in intimate partner violence increasingly considers underlying dynamics rather than one-off events.[20] Moreover, the Draft AI Act explicitly excludes systems where distortion or harm arises from dynamics of the user-base entwined with an AI system,[21] excluding salient areas such as discriminatory ratings or recommendations on dating apps and online markets.[22]

13  Furthermore, an AI system may classify people (e.g. emotionally), while a *separate* downstream actor uses that classification harmfully.[23] How and to whom should the Draft AI Act's prohibition apply? Upstream classification with both useful and harmful potency is a difficult-to-govern 'dual use' artefact familiar in technology policy.[24] Yet digital intermediaries frequently benefit from a mix of illegal and legal activity, for example in advertising or copyright.[25] The Draft AI Act does not rise to this challenge.

### b) Comparison to Existing Union Law

14  Even where these prohibitions apply, they add little to existing EU law. Both resemble the Unfair Commercial Practices Directive, which prohibits commercial practices if they 'materially [distort] or [are] likely to materially distort the economic behaviour with regard to the product of the average consumer [...] or of the average member of the group'.[26] In that Directive, the latter condition is triggered if a vulnerability on the basis of 'physical infirmity, age or credulity' is foreseeable.[27] Commercial practices are broad, including advertising, communications and other 'acts' relating to goods or services.

15  The importance of the Draft AI Act's expansion beyond the Unfair Commercial Practices Directive to *non-economic* decision-making[28] is limited by the Draft AI Act's harm requirements. Legislators may wish to note that workable alternatives exist to harm tests in information law, such as 'reasonable per-

---

16  See, from DG CONNECT, Gabriele Mazzini, 'A European Strategy for Artificial Intelligence' (*2nd ELLIS Workshop in Human-Centric Machine Learning (YouTube recording)*, 10 May 2021) https://youtu.be/OZtuVKW qhl0?t=10346 accessed 22 June 2021, at 2:52:26 *et seq.*

17  Marijn Sax, 'Between Empowerment and Manipulation: The Ethics and Regulation of For-Profit Health Apps' (PhD Thesis, Univeristeit van Amsterdam (UvA) 2021) 110–12.

18  See e.g. Oscar H Gandy, *Coming to Terms with Chance: Engaging Rational Discrimination and Cumulative Disadvantage* (Routledge 2009).

19  See generally Nick Seaver, 'Captivating Algorithms: Recommender Systems as Traps' (2019) 24 Journal of Material Culture 421. Harms are identified especially in relation to children, see e.g. Beeban Kidron and others, 'The Cost of Persuasive Design' (*5 Rights Foundation*, June 2018) https://5rightsfoundation.com/uploads/5rights-disrupted-childhood-digital-version.pdf.

20  See generally Evan Stark and Marianne Hester, 'Coercive Control: Update and Review' (2019) 25 Violence Against Women 81 (on how the concept of *coercive control* entered English law due to how 'discrete, injurious assaults [were] too narrow to capture [patterns] of coercion').

21  AI Act, recital 16 ('intention may not be presumed if the distortion of human behaviour results from factors external to the AI system which are outside of the control of the provider or the user').

22  Jevan A Hutson and others, 'Debiasing Desire: Addressing Bias & Discrimination on Intimate Platforms' (2018) 2 Proc ACM Hum-Comput Interact 73:1; Karen Levy and Solon Barocas, 'Designing against Discrimination in Online Markets' (2017) 32 Berkeley Tech LJ 1183.

23  See, on multi-stage profiling, Reuben Binns and Michael Veale, 'Is That Your Final Decision? Multi-Stage Profiling, Selective Effects and Article 22 of the GDPR' (Presented at PLSC-EU 2019, on file with authors 2021). Classification itself may be considered a form of harm (of representation), but there is little legal protection around this. See Reuben Binns, 'Fairness in Machine Learning: Lessons from Political Philosophy' (2018) Conference on Fairness, Accountability and Transparency (FAT* 2018), 8.

24  See John Forge, 'A Note on the Definition of "Dual Use"' (2010) 16 Sci Eng Ethics 111.

25  See e.g. coordinating intermediaries in real-time bidding, Michael Veale and Frederik Zuiderveen Borgesius, 'Adtech and Real-Time Bidding under European Data Protection Law' [2021] German Law Journal; Cristiana Santos and others, 'Consent Management Platforms Under the GDPR: Processors and/or Controllers?' in *Privacy Technologies and Policy* (Cham, Nils Gruschka and others eds, Springer International Publishing 2021).

26  Directive 2005/29/EC of the European Parliament and of the Council of 11 May 2005 concerning unfair business-to-consumer commercial practices in the internal market and amending Council Directive 84/450/EEC, Directives 97/7/EC, 98/27/EC and 2002/65/EC of the European Parliament and of the Council and Regulation (EC) No 2006/2004 of the European Parliament and of the Council ('Unfair Commercial Practices Directive'), OJ L 149/22, art 5.

27  Directive 2005/29/EC of the European Parliament and of the Council of 11 May 2005 concerning unfair business-to-consumer commercial practices in the internal market and amending Council Directive 84/450/EEC, Directives 97/7/EC, 98/27/EC and 2002/65/EC of the European Parliament and of the Council and Regulation (EC) No 2006/2004 of the European Parliament and of the Council ('Unfair Commercial Practices Directive'), OJ L 149/22, art 5.

28  The limits of the UCPD to transactional decisions relate to the desire to have both a harmonised yet open-ended definition of fairness, which would not be a mechanism for the Member States with a history of moral standards in consumer law to reintroduce them through creative interpretation and create barriers to European trade. See generally Hans-Wolfgang Micklitz, 'Unfair Commercial Practices and Misleading Advertising' in Hans-Wolfgang Micklitz and others (eds), *Understanding EU Consumer Law* (Intersentia 2009).

son' requirements creating flexible red lines, which may strike a fairer balance in these complex situations.[29]

### c) Ineffectiveness

16 Lastly, unlike the Unfair Commercial Practices Directive which focusses on use, the Draft AI Act also prohibits the *sale* of in-scope manipulative systems, e.g. to oppressive regimes. Yet vendors can attempt to dodge this requirement by selling general purpose AI systems which can be (re)configured by a user. The recitals indicate that the manipulation provisions relate to systems '*intended* to distort human behaviour' [emphasis added].[30] Few vendors would admit to such intention. Disguising the 'true' market for digital products is already common practice – consider stalkerware disguised as child trackers.[31] Reconfiguration (or to use the industry term, 'democratisation') of AI is a significant trend, typified by *AI-as-a-service*.[32]

17 In sum, the prohibitions concerning manipulative AI systems may have little practical impact.

## 2. Social Scoring

18 The second grouping of prohibitions relate to concerns about so-called 'social scoring'. The Draft AI Act prohibits the sale or use of AI systems i) used by or on behalf of public authorities, ii) to generate 'trustworthiness' scores and which ii) lead to either unjustified or disproportionate treatment of individuals or groups, or detrimental treatment which, while justifiable and proportionate, occurs in an unrelated 'context' from the input data.

19 Trustworthiness is not defined in the Draft AI Act, but can be understood as a combination of attributes that indicate that an entity will not betray another due to bad faith such as misaligned incentives, lack of care, disregard for promise-keeping (*commitment*) or through ineptitude at a task (*competence*).[33] Understood in this way, many scoring practices are in-scope.[34]

### a) Scope of Same Context

20 This 'same-context' exemption in ii) appears designed to keep reputation systems out of scope, and recalls theoretical work in privacy, including *contextual integrity*.[35] Yet the exemption will be difficult to operationalise.

21 It is unclear whether the citizen scoring characterising the 'datafied welfare state', commonly built with broad private sector datasets augmenting administrative data, will be in-scope.[36] If context is viewed narrowly, scoring can only relate to input data concerning interactions with public authorities. A wider view however might consider credit card records and welfare support as the same context, as both involve financial flows. The Commission anticipates that a system which 'identifies at-risk children in need of social care' would be out of context if 'based on insignificant or irrelevant social 'misbehaviour' of parents, e.g. missing a doctor's appointment or divorce'.[37] The 'European Artificial Intelligence Board' may end up with the job of clarifying, but its guidance is only advisory.[38]

22 Public employment may also be impacted. Automated 'social media background checks' to score online lives seemingly concern both *competence* and *commitment* aspects of trustworthiness.[39] Where such systems risk detrimental outcomes, such checks would likely be prohibited due to a contextual disconnect. A public sector body using a 'trustworthiness'–related ranking of freelancers provided to all by *LinkedIn* creates a further set of questions. Would LinkedIn also be liable for providing such a service to the public sector?

### b) Allocation of Responsibility

23 What counts as an AI system 'leading to' an outcome is also unclear. Vendors will disavow negative outcomes and blame them on their users. Users (who are the authorities, rather than the citizens[40]) will claim that scoring was never a fully determinative factor. The result appears to be no entity clearly liable at all.

24 The logic behind restricting this prohibition to the public sector remains unclear. So-called AI firms control crucial infrastructures, such as delivery, telecommunications or transport. Exclusion can bring individuals grave socioeconomic consequences similar to the exclusion of state-provided services.[41] As with manipulation, the EU legislator has some work to do to make this provision clearly applicable to anything.

---

29 For example, in Canada the federal private sector privacy law PIPEDA utilises flexible red lines. It stipulates that an 'organization may collect, use or disclose personal information only for purposes that a reasonable person would consider are appropriate in the circumstances' – a provision the regulator describes as establishing 'no-go zones'. See Personal Information Protection and Electronic Documents Act 2000 (Canada) s 5 (3).

30 AI Act, recital 16. Intention gets three mentions in this recital alone.

31 Diarmaid Harkin and others, 'The Commodification of Mobile Phone Surveillance: An Analysis of the Consumer Spyware Industry' (2020) 16 Crime, Media, Culture 33.

32 Jennifer Cobbe and Jatinder Singh, 'Artificial Intelligence as a Service: Legal Responsibilities, Liabilities, and Policy Challenges' (*Preprint available on SSRN*, 4 December 2021) https://ssrn.com/abstract=3824736 accessed 7 June 2021. The industry preferred term 'democratisation' is discussed in Sudhir Hasbe and Ryan Lippert, 'Democratization of Machine Learning and Artificial Intelligence with Google Cloud' (*Google Cloud Blog*, 16 November 2020) https://perma.cc/DL86-RJW3 accessed 4 May 2021.

33 Margaret Levi and Laura Stoker, 'Political Trust and Trustworthiness' (2000) 3 Annual Review of Political Science 475.

34 This structure differs from the HLEG-AI's initial recommendation in this area, to prohibit 'mass scale scoring' assessing 'moral personality' or 'ethical integrity'. See High-Level Expert Group on Artificial Intelligence, 'Ethics Guidelines for Trustworthy AI' (April 2019) 34; High-Level Expert Group on Artificial Intelligence, 'Policy and Investment Recommendations for Trustworthy AI' (26 June 2019) 20.

35 See generally Helen Nissenbaum, *Privacy in Context: Technology, Policy, and the Integrity of Social Life* (Stanford University Press 2010).

36 For example, the Mosaic dataset of Experian, a consumer credit reporting company, encompasses a 'broad and accurate range of demographic, socio-economic and behavioural characteristics on each adult and household. See Lina Dencik and others, 'Data Scores as Governance: Investigating Uses of Citizen Scoring in Public Services' (*Data Justice Lab, Cardiff University*, 2018) 92–93 https://perma.cc/39CY-H8L7 accessed 21 August 2020; see generally across the EU, Algorithm Watch, 'Automating Society Report 2020' (October 2020) https://automatingsociety.algorithmwatch.org accessed 20 June 2021.

37 See, from DG CONNECT, Mazzini (n 16).

38 AI Act, art 58(c).

39 Miranda Bogen and Aaron Rieke, *Help Wanted – An Exploration of Hiring Algorithms, Equity and Bias* (Upturn 2018) 38–39.

40 AI Act, art 3(4).

41 K Sabeel Rahman, 'The New Utilities: Private Power, Social Infrastructure, and the Revival of the Public Utility Concept' (2017–18) 39 Cardozo L Rev 1621.

## 3. Biometric Systems

25  The Draft AI Act bans some *uses* of 'real-time' biometric systems in publicly accessible spaces by law enforcement. An example of such a system would be a large-scale CCTV network coupled with facial recognition software. Law enforcement use of biometric identification is regulated in the Law Enforcement Directive,[42] which is the GDPR-type instrument for the police and similar.[43] Systems such as facial recognition have been easier to authorise for law enforcement purposes than other uses, such as that for a company's interest, which typically fall under the GDPR. The proposed strengthening would make the AI Act *lex specialis* to the Law Enforcement Directive, with this provision based upon TFEU Article 16 rather than 114 as the rest of the Draft AI Act is.[44]

26  The Draft AI Act enables Member States to authorise certain uses that fall within an exhaustive list of exceptions if accompanied by certain safeguards. Roughly summarised, the exemptions are:

- a 'targeted search for specific potential victims of crime, including missing children';

- the 'prevention of a specific, substantial and imminent threat to the life or physical safety of natural persons or of a terrorist attack'; and

- the 'detection, localisation, identification or prosecution' of a perpetrator or suspect of a crime with a maximum sentence of at least 3 years that would allow for the issuing of a European Arrest Warrant.

### a) Three Shortcomings

27  We can observe, firstly, that unlike the above prohibitions, this provision would allow such biometric systems to be 'placed on the market', meaning EU vendors can sell biometric systems which would be illegal to use in the EU to oppressive regimes in third countries.[45] Examples of such practices are the French firm Idemia/Morpho selling facial recognition to the Shanghai Public Security Bureau, or the Dutch firm Noldus selling facial expression analysis tool 'FaceReader' to the Chinese Ministry of Public Security.[46]

28  Secondly, only 'real-time' systems that capture, compare, and identify 'instantaneously, near-instantaneously or in any event without a significant delay' are prohibited. This excludes 'post' systems which, for example, biometrically analyse footage after an event, for example to identify individuals at protests after-the-fact,[47] and systems that categorise individuals biometrically.[48] As online spaces are also out-of-scope, live biometric identification on e.g. video streams is also excluded.[49]

29  Thirdly, the prohibition does not ban actors from using remote biometric identification for non-law enforcement purposes, such as crowd control or public health. These uses typically fall under the GDPR. Roughly summarised, in the absence of a proportionate Member State law authorising such biometrics, the GDPR places a requirement of high-quality, individual consent for each scanned person which is effectively impossible to fulfil.[50]

### b) Need for Pre-Authorisation of "Individual Use"

30  This provision also introduces pre-authorisation familiar from state surveillance law. Competent authorities' 'individual use' of a biometric system must be pre-authorised by a judicial authority or independent administrative authority (or in an emergency, shortly afterwards).[51] Analogous CJEU case-law regarding data retention indicates authorising bodies must have a 'neutral stance', notably excluding public prosecutors.[52] The Draft AI Act also requires the decision of this body to be final,

---

42  Directive (EU) 2016/680 of the European Parliament and of the Council of 27 April 2016 on the protection of natural persons with regard to the processing of personal data by competent authorities for the purposes of the prevention, investigation, detection or prosecution of criminal offences or the execution of criminal penalties, and on the free movement of such data, and repealing Council Framework Decision 2008/977/JHA, OJ L 119/89 ('Law Enforcement Directive').

43  Regulation (EU) 2016/679 of the European Parliament and of the Council of 27 April 2016 on the protection of natural persons with regard to the processing of personal data and on the free movement of such data, and repealing Directive 95/46/EC (General Data Protection Regulation) OJ L 119/1.

44  AI Act, recital 23; see further Theodore Christakis and Mathias Becuywe, 'Pre-Market Requirements, Prior Authorisation and Lex Specialis: Novelties and Logic in the Facial Recognition-Related Provisions of the Draft AI Regulation' (*European Law Blog*, 5 April 2021) https://europeanlawblog.eu/2021/05/04/pre-market-requirements-prior-authorisation-and-lex-specialis-novelties-and-logic-in-the-facial-recognition-related-provisions-of-the-draft-ai-regulation/ accessed 6 May 2021. National courts have not interpreted the LED as prohibiting facial recognition in its entirety, see e.g. *R (on the Application of Bridges) v South Wales Police* [2020] EWCA Civ 1058 (a case relating to pre-Brexit facts).

45  Some of these sales may be regulated or require transparency or authorisation under Regulation (EU) 2021/821 of the European Parliament and of the Council of 20 May 2021 setting up a Union regime for the control of exports, brokering, technical assistance, transit and transfer of dual-use items (recast), OJ L 206/1 ('Dual Use Regulation'); which has been criticised by civil society groups in relation to a lack of prohibitions, see Access Now and others, 'New EU Dual Use Regulation Agreement "a Missed Opportunity"' (25 March 2021) https://perma.cc/P49G-P3ZR accessed 21 June 2021.

46  Amnesty International, 'Out of Control: Failing EU Laws for Digital Surveillance Export' (September 2020) https://perma.cc/2GU5-84ZT accessed 21 June 2021.

47  See further European Data Protection Board and European Data Protection Supervisor, 'Joint Opinion 5/2021 on the Proposal for a Regulation of the European Parliament and of the Council Laying down Harmonised Rules on Artificial Intelligence (Artificial Intelligence Act)' (18 June 2021) para 31.These are however included in the Title III regime, discussed below.

48  Categorisation systems fall under Title IV and have weak transparency requirements, but even these have law enforcement exceptions. They could be generically added to Title III (Annex II) under delegated legislation, as the area includes biometric categorisation; regarding law enforcement it may fall under Annex II, paras 6(f–g).

49  AI Act, recital 9.

50  These are the most relevant conditions in GDPR, art 9; others may apply but only in extremely unusual situations. See generally European Data Protection Board, 'Guidelines 3/2019 on Processing of Personal Data through Video Devices (Version 2.0)' (*EDPB*, 29 January 2020) https://edpb.europa.eu/our-work-tools/our-documents/guidelines/guidelines-32019-processing-personal-data-through-video_en accessed 6 May 2021.

51  Such a system resembles national intelligence structures and the role of, for example, the *Toetsingscommissie Inzet Bevoegdheden* in the Netherlands, or the Investigatory Powers Commissioner in the United Kingdom.

52  Case C-746/18 *HK v Prokuratuur* ECLI:EU:C:2021:152 [54]. The French *parquet* is an example of a body that currently authorises such surveillance but may not be allowed to in these cases.

whereas in some Member States the executive can ignore similar bodies.[53]

31 What constitutes an 'individual use' to be authorised is unclear. In signals intelligence, controversial warrants can be *thematic*, relating to broad organisations, places or purposes.[54] In the Draft AI Act, it is unclear if 'individual' could be an individual *purpose*, e.g. authorising biometrics relating to all those on a missing children list or subject to a European Arrest Warrant. As the Draft AI Act does not explicitly require transparency over the number and type of authorisations issued, public scrutiny may be challenging.[55]

32 Either way, any authorisation of biometrics necessitates installing re-purposable infrastructure. Many already argue the Draft AI Act legitimises rather than prohibits population-scale surveillance. The European Data Protection Board (EDPB) and the European Data Protection Supervisor (EDPS) 'call for a general ban on any use of AI for an automated recognition of human features in publicly accessible spaces […] in any context'.[56] Many NGOs have come out similarly.[57]

### III. Title III Regime: High-Risk Systems

33 The Title III regime governs AI system that pose 'high-risk' to 'health, safety and fundamental rights'[58] in number of defined applications, products and sectors. The regime is based on and entwined with the *New Legislative Framework* (NLF) (the *New Approach* when introduced in 1985), a common EU approach to the regulation of certain products such as lifts, medical devices, personal protective equipment and toys.[59]

#### 1. Scope

34 While the Draft AI Act as a whole applies to all 'AI systems', Title III, on high-risk AI systems, only applies to two sub-categories of AI systems:

35 Firstly, AI systems that are products or safety components (broadly construed) of products already covered by certain Union health and safety harmonisation legislation (such as toys, machinery, lifts, or medical devices).[60]

36 Secondly, 'standalone' AI systems specified in an annex for use in eight fixed areas:[61]

- biometric identification and categorisation (both 'remote', as in Title II above, and applied 'post' the event);
- management and operation of critical infrastructure;
- educational and vocational training;
- employment, worker management and access to self-employment;
- access to and enjoyment of essential services and benefits;
- law enforcement;
- migration, asylum and border management;
- administration of justice and democracy.

37 The Commission can, subject to Parliament or Council veto, add *sub-areas* within these areas if the application poses similar risk to an existing in-scope application, but cannot add new areas entirely.[62]

### 2. The Draft AI Act in the Context of the New Legislative Framework (NLF)

38 Under NLF regimes, a manufacturer must undertake pre-marketing controls undertaken to establish products' safety and performance, through *conformity assessment* to certain *essential requirements* laid out in law. Manufacturers then mark conforming products with 'CE'; marked products enjoy EU freedom of movement. The philosophy of the NLF is that '[t]he manufacturer, having detailed knowledge of the design and production process, is best placed to carry out the complete conformity assessment procedure. Conformity assessment should therefore remain the obligation of the manufacturer alone.'[63] This distinguishes NLF regimes (including the Draft AI Act) from pharmaceutical regulation, where a public authority (e.g. the European Medicines Agency) carries out an assessment *themselves* before granting pre-marketing approval.[64]

### 3. Essential Requirements and Obligations

39 The Act contains an extensive list of essential requirements (Chapter 2) which connects to obligations of regulated actors (Chapter 3). The vast majority of all obligations fall on the 'provider: in short, person or body that develops an AI system or that has an AI system developed with a view to placing it on

---

53 Such as the French *Commission Nationale de Contrôle des Techniques de Renseignement*, which can express disapproval but not overrule the Prime Minister.

54 See e.g. those avowed by the UK in the atmosphere of post-Snowden scrutiny at Intelligence and Security Committee of Parliament, *Privacy and Security: A Modern and Transparent Legal Framework* (2015) 111.

55 As required, for example, by the UK's Investigatory Powers Act 2016, s 234(2)(d). Transparency of surveillance regimes differ across countries, see e.g. in the Dutch context Quirine Eijkman and others, 'Dutch National Security Reform Under Review: Sufficient Checks and Balances in the Intelligence and Security Services Act 2017?' (*IViR, University of Amsterdam*, March 2018) 40–41 https://perma.cc/LJ4Y-ZQRQ accessed 21 June 2021.

56 European Data Protection Board and European Data Protection Supervisor (n 47).

57 For example, over 60 NGOs are running a campaign 'Reclaim Your Face' at https://reclaimyourface.eu.

58 AI Act, recital 43, art 7(2).

59 Little is actually new about the NLF. See Harm Schepel, *The Constitution of Private Governance: Product Standards in the Regulation of Integrating Markets* (Hart 2005) 64 (stating '[t]he 'New Approach' will most likely stay 'new' forever, but was not so 'new' even when it was launched').

60 See the list in AI Act, Annex II. Section A lists other 'New Approach' legislation; section B legislation is older-style product safety legislation (with a stronger role for public bodies and more detailed requirements in law) which are instead amended by Title XII to introduce new AI Act–related considerations for future delegated acts in those areas.

61 AI Act, Annex III.

62 AI Act, arts 7, 73.

63 Decision No 768/2008/EC of the European Parliament and of the Council of 9 July 2008 on a common framework for the marketing of products, and repealing Council Decision 93/465/EEC OJ L 218/82.

64 Note however that the AI Act is unusual in proposing that for relevant biometric systems intended to be put into service by law enforcement, immigration or asylum authorities or EU institutions, conformity assessment does require a public body, who in practice will be a data protection authority, supervising agency of those authorities, or the EDPS. See AI Act, arts 43(1), 63(5).

the market or putting it into service under its own name or trademark.[65]

40 Providers of high-risk AI systems must create a *quality management system*,[66] a standardised practice already widely present in firms.[67] The Draft AI Act specifies what this entails, featuring a documented *risk management system* updated throughout the system's lifetime.[68]

41 Datasets to train AI systems must meet *data quality criteria*, including in relation to relevance, representativeness, accuracy, completeness, and application-area specific properties. Despite some requirements seeming steep – datasets being 'free of errors and complete'[69], which they often are far from[70] – datasets only need to meet these potentially steep requirements 'sufficiently' and 'in view of the intended purpose of the system'.[71]

42 Much attention has been paid to the potential for AI systems to facilitate indirect discrimination, in principle illegal under EU law.[72] It is difficult to detect this potential unless providers know the relevant protected (and often sensitive) characteristics of affected individuals and communities.[73] However, the GDPR restricts the use of ethnicity data and similar sensitive data, with no specific EU-level exemption for bias detection.[74] The Draft AI Act provides such an exemption.[75] The exemption can only be used in relation to high-risk systems, and only by those systems' providers. This leaves non-high-risk providers unable to rely on it.[76] The exemption does not provide a route for upstream data brokers to collect sensitive data on others' behalf, or to later sell to high-risk providers.

43 The Draft AI Act contains obligations concerning the *accuracy, robustness and cybersecurity* of systems themselves, with particular regard to discrimination as systems learn,[77] and adversarial machine learning.[78] There is no explicit discussion of leakage of training data or other personal data from models.[79]

44 Providers must create *technical documentation* in line with a (Commission–amendable) Annex. The requirements are extensive; we refer the reader to them. The provider does not have to publish the technical documentation or provide it except to organisations involved in regulation or conformity assessment. However, separate provisions indicate what information must be provided as a form of *user transparency*, and what information must be *registered in a public database*. In Table 1, we abstract and group the (most salient) categories of information to be provided.

45 Providers must facilitate *logging* to allow traceability appropriate to a system's risks. For biometric systems, logging must include periods of use; the reference database used; and any input data leading to a match. Providers must implement a mechanism to record the identities of the 'two natural persons' checking a biometric match before it is used, and instruct a user to only use it with such a check.[80] Providers must only keep logs (for an appropriate amount of time) 'to the extent such logs are under their control',[81] else the user must instead.[82]

46 Providers must build for *human oversight*, incorporating 'human-machine interface tools' to ensure systems 'can be effectively overseen by natural persons'.[83] In data protection law, human oversight typically relates to human dignity.[84] In the AI Act, human oversight instead relates to minimising risks to health, safety and fundamental rights.[85] A 'four-eyes' principle requires biometric identification systems to be designed so that two natural persons can sign off on any identification and have their identities logged, and for instructions to specify that they must.[86]

| | Public | Users | Documentation |
|---|---|---|---|
| identity; contact details | | (available publicly) | (assumed) |
| member states in use | | | (available publicly) |
| purpose | | | |
| conformity assessment information | | | |
| relevant standards | | | |
| instructions for use | | | |
| human oversight & technologies | | | |
| accuracy, robustness, cybersecurity | "level of"; metrics (accuracy) | | metrics; test logs; test reports |
| risky use circumstances | | | "detailed information" |
| performance on persons/groups | | | "detailed information" |
| input data | | "where appropriate, specifications" | datasheets incl. training datasets and main characteristics; provenance; labelling procedures; data cleaning |
| pre-determined changes | | | "detailed description"; techniques to ensure "continuous compliance" |
| lifecycle information | | expected lifetime; maintenance info | "description of any change made to the system" |
| post-market monitoring | | | "detailed description [of plan]" |
| risk management system | | | "detailed description" |
| design specifications | | | "general logic"; key choices and assumptions; optimisation function; trade-off decisions; description of hardware and interacting systems |
| methods and steps of development | | | role of pre-trained models/tools; computational resources used; training methodologies |

**Table 1:** Main categories of information provided (white) (or partially (grey), or not (black)) to the public, to users, and kept by providers in technical documentation. Not fully exhaustive and grouped for comparison; refer to the Act for full information.

47 Somewhat strangely, no obligations for human oversight flow directly from the Act to a user. In relation to human oversight, users must simply follow the instruction manual. If a provider has a high regulatory risk appetite, their instructions may not mandate rigorous oversight, yet the user would not be liable.[87] Nevertheless, this does extend some liability to the user, also in relation to monitoring systems during use,[88] and affected individuals and groups may be able to draw upon these obligations in national tort actions.

48 Interestingly, a leaked version of the Draft AI Act required providers to specify *organisational measures*, notably similar to data protection guidelines,[89] including to ensure that overseers 'can decide not to use the high-risk AI system or its outputs in any particular situation without any reason to fear negative consequences', and obliged users to follow these.[90] In a *volte-face*, the final Proposal instead emphasises the 'user's discretion in organising its own resources and activities for the purpose of implementing the human oversight measures indicated by the provider.'[91] Statements about the need for 'competence, training and authority' only make the recitals.[92]

### 4. Conformity Assessment and Presumption

49 These requirements are applied to providers as they must undergo *conformity assessment*. To understand conformity assessment on-the-ground, we need to explain two other actors in the Draft AI Act: standardisation organisations and notified bodies.

### a) Harmonised Standards & European Standardisation Organisations

50 Arguably the most important actors in the Draft AI Act are the double-act of CEN (European Committee for Standardisation) and CENELEC (European Committee for Electrotechnical Standardisation). This may surprise readers; neither are mentioned in the text. These are two of three European Standardisation Organisations (ESOs) that the Commission can mandate to develop *harmonised standards*.[93]

---

87 AI Act, art 29(1).
88 AI Act, art 29(4).
89 See Article 29 Working Party, 'Guidelines on Automated Individual Decision-Making and Profiling for the Purposes of Regulation 2016/679 (WP251rev.01)' (6 February 2018) (stating that those overseeing the decisions must have the 'authority and competence' to do so); see Michael Veale and Lilian Edwards, 'Clarity, Surprises, and Further Questions in the Article 29 Working Party Draft Guidance on Automated Decision-Making and Profiling' (2018) 34 Computer Law & Security Review 398, 401 (on why this is an organisational matter).
90 Leaked AI Act, arts 11(3)(e), 18(2). The leak, dated in January, was first made available by policy subscription service POLITICO Pro (link unavailable), and republished in Natasha Lomas, 'EU Plan for Risk-Based AI Rules to Set Fines as High as 4% of Global Turnover, per Leaked Draft' (*TechCrunch*, 14 April 2021) https://techcrunch.com/2021/04/14/eu-plan-for-risk-based-ai-rules-to-set-fines-as-high-as-4-of-global-turnover-per-leaked-draft/ accessed 1 July 2021.
91 AI Act, art 29(2).
92 AI Act, recital 48.
93 Regulation (EU) No 1025/2012 of the European Parliament and of the Council of 25 October 2012 on European standardisation, amending

51 Following a mandate, if these organisations adopt a standard relating to the Draft AI Act, providers can follow this standard, rather than interpreting the *essential requirements.* If following the standard, providers enjoy a presumption of conformity.[94]

52 Standards can cover a legal instrument's entire scope, or only specialist areas. For instance, the essential requirements of the Toy Directive for trampolines can be fulfilled through EN 71-14:2018; kids' chemistry sets through EN 71-4:2013; and 'olfactory board games, gustative games, and cosmetic kits' through EN 71-13:2014. Standards are not free – copyright is owned by national standards bodies, and each usually costs a few hundred Euros to purchase. The Commission anticipates that the Draft AI Act standards (it is not clear if general or specific) will first appear in the EU's Official Journal in 2024–2025, aligned with when the Draft AI Act would be applicable.[95] After the industrial lobbying common in standards bodies, some aspects may look quite different from the essential requirements.[96]

53 In theory, providers do not have to follow such harmonised standards. Instead, providers could interpret the Draft AI Act's essential requirements for themselves.[97] This is easier said than done. Harmonised standards are both cheaper for producers, and a safer bet.[98] They are not as voluntary as the Commission argues. Essential requirements are often not realistically suitable for direct application.[99] Harmonised standards often function as a necessary point of reference for compliance through essential requirements.[100] In the Draft AI Act, the requirement to consult harmonised standards is explicit.[101] Consequently, standardisation is arguably where the real rule-making in the Draft AI Act will occur.

### b) Controversies of Harmonised Standards

54 The practice of delegating rule-making to bodies governed by private law such as CEN/CENELEC is controversial and sits on increasingly shaky legal ground.

55 Firstly, outside the field of AI regulation, it has long been argued that 'there are structural reasons why the [New Legislative Framework] might serve the European consumer ill'.[102] Under-resourced consumer organisations struggle to participate in arcane private standardisation processes,[103] yet the outputs are important standards Member States *must* recognise. In the case of the Draft AI Act many rights and freedoms are at stake. It is unclear whether limited existing efforts to include stakeholder representation will enable the deep and meaningful engagement needed from affected communities.[104] The vast majority will have absolutely no experience of standardisation, and may lack EU-level representation.[105] Moreover, the European Parliament has no binding veto over harmonised standards mandated by the Commission.[106]

56 Secondly, the Draft AI Act's value-laden nature might plant a constitutional bomb under the New Legislative Framework. Even 'technical' safety standards entail value-laden choices about, for example, thresholds of acceptable risk, taken under uncertainty.[107] The CJEU appears to be slowly recognising private standardisation bodies mandated as *de facto* NLF rule-makers cannot be free from judicial scrutiny.[108] Yet the NLF constitutionally relies on them being so. If such standardisation bodies are not free from judicial scrutiny, the NLF model of harmonised standards risks classification as unlawful delegation of the Commission's rulemaking power to private bodies.[109] Its novel incorporation of broad fundamental rights topics into the NLF make the Draft AI Act spotlight this tension of legitimacy.[110]

---

57 In sum, the Commission's long practice of privately outsourcing complex negotiations has been controversial for years. The Draft AI Act may trigger more attention to this constitutional problem.[111]

### c) Self-Assessment and the (Limited) Role of Notified Bodies

58 For some products, NLF-regulated manufacturers can affix a CE certificate after 'conformity assessment based on internal control'. In the Draft AI Act this means that they self-assess that their *quality management system*, *system-specific technical documentation*, and *post-market monitoring plan* follow either the essential requirements or a relevant harmonised standard/common specification.

59 However, under some conditions, NLF self-assessments require approval by an independent technical organisation of the provider's choosing known as a *notified body*. Notified bodies are typically private sector certification firms. They are accredited by Member States' *notifying authorities*.[112] Examples of notified bodies range from giants such as the German and Austrian TÜV groups, multinationals with thousands of employees who inspect and audit in a huge number of sectors, to more specialist bodies such as the Dutch *Liftinstituut*, which certifies elevators. In theory, notified bodies are transparently listed online and subject to organisational standards.[113] In practice, little is known about their activities, particularly due to frequent outsourcing.[114]

60 Despite pages of the Draft AI Act establishing a regime for AI Act-specific notified bodies, there are almost no situations where their services are required. For most standalone high-risk systems (and eventually, all such systems), providers can mark the systems as in conformity using only self-assessment.

61 Only listed high-risk applications within the area of 'biometric identification and categorisation of natural persons' must use AI Act-specific notified bodies – (initially only remote identification systems). Once harmonised standards or common specifications covering those systems exist, only self-assessment is needed.[115] As the Commission hopes harmonised standards will exist before the application of the Regulation,[116] AI Act-specific notified bodies may indeed *never* be required, even for biometric systems.

62 AI products or components that fall under other in-scope harmonisation instruments, such as medical devices, may also require notified bodies created under the respective regime. This applies only if the product usually requires a notified body for conformity, as not to create a loophole where AI-powered products could self-assess whereas other products could not.

63 Political science has shown how regulatory intermediaries such as notified bodies play important roles beyond assurance, for example in translating rules, providing know-how to targets of regulation, and providing feedback to regulators and standard-setters.[117] The Draft AI Act obliges notified bodies to participate in co-ordination activities.[118] However, as AI Act specific notified bodies may never exist, at least in relation to non-biometric applications, this obligation seems a little futile, and leaves big gaps in knowledge flows regarding how the Draft AI Act is functioning on-the-ground.

64 In sum, the Draft AI Act gives a large role to two private standardisation organisations. CEN and CENELEC can adopt standards relating to the Draft AI Act; standards that AI providers will follow in practice. Notified bodies checking a provider's self-assessment may play a small role, but there are few situations where they are required.

65 We will come back to the issue of enforcement and oversight of conformity assessment in section IV.3. (para. 76 et seq.) below, as this cuts across all levels of risk. For now, we turn to the 'limited risk' group of Title IV.

## IV. Title IV: Specific Transparency Obligations

66 Title IV lays out three transparency obligations – two for AI users, one for AI providers – that apply to all AI systems that meet their criteria.[119] The Commission has no powers to alter Title IV.

### 1. 'Bot' Disclosure

67 Providers of AI systems intended to interact with natural persons (hereafter 'bots' for short[120]) must design their systems such that individuals are informed they are interacting with a bot, unless it would be contextually obvious that individuals are interacting with a bot, or if the bot use is authorised by law to prevent criminal offences.[121]

68 Bot disclosure laws are not new, although none are quite like this one. In 2018, California passed the *Bolstering Online Transparency* (BOT) *Act*.[122] The BOT Act targets individuals, making it unlawful for any person to use a bot[123] to interact online with a Californian intending to mislead them about its artificial identity to incentivise a purchase or influence an electoral vote without clear, conspicuous disclosure.

---

111 Schepel (n 99) 192.
112 Jean-Pierre Galland, 'The Difficulties of Regulating Markets and Risks in Europe through Notified Bodies' (2013) 4 Eur J Risk Reg 365, 368–69.
113 Notified bodies are published in the Official Journal and on the EU's online NANDO (New Approach Notified and Designated Organisations) database. https://ec.europa.eu/growth/tools-databases/nando. They follow organisational standards including EN ISO/IEC 17000, and varying legal obligations.
114 Galland (n 112) 369.
115 AI Act, art 43(1).
116 European Commission, 'AI Act Impact Assessment' (n 95) 57.
117 Kenneth W Abbott and others, 'Theorizing Regulatory Intermediaries: The RIT Model' (2017) 670 The ANNALS of the American Academy of Political and Social Science 14; Kira JM Matus and Michael Veale, 'Certification Systems for Machine Learning: Lessons from Sustainability' [2021] Regulation & Governance. Note that such feedback can be useful but can also be geared towards increasing the profitability of notified bodies by reducing audit costs and rigour. See Jean-Pierre Galland, 'Big Third-Party Certifiers and the Construction of Transnational Regulation' (2017) 670 The ANNALS of the American Academy of Political and Social Science 263, 274.
118 AI Act, art 33(11).
119 AI Act, art 52.
120 While we use this as convenient shorthand, the term is fraught with definitional challenges; see generally Robert Gorwa and Douglas Guilbeault, 'Unpacking the Social Media Bot: A Typology to Guide Research and Policy' (2020) 12 Policy & Internet 225.
121 AI Act, art 52(1). Note that the criminal prevention exemption does not apply to systems that help reporting of crime.
122 California's Business & Professions Code §17940, *et seq*. It has been in force since July 2019. A proposed federal bill, the 'Bot Disclosure and Accountability Act', died in the 2019–21 Congress.
123 Defined as 'an automated online account where all or substantially all of the actions or posts of that account are not the result of a person'.

69 The European Commission's voluntary Code of Practice on Disinformation commits signatory platforms to '[e]stablish clear marking systems and rules for bots and ensure their activities cannot be confused with human interactions'.[124] A strengthened version is expected in Autumn 2021, linking with the Draft AI Act to tackle broad 'inauthentic behaviour'[125] and potentially becoming a code of conduct under the proposed Digital Services Act.[126]

70 In the Draft AI Act, bot disclosure liability flows to *providers*, not users or platforms. This somewhat resolves the objections of scholars who criticise bot disclosure laws and proposals because practical enforcement may force exposure of the natural person behind allegedly automated content.[127] Enforcement of the Draft AI Act does not require unmasking the users; i.e. the person or body using a bot. Instead, the Draft AI Act identifies technology providers through surveilling the market for products. *Market surveillance authorities* have powers to compel online intermediaries to help them, but only intermediaries facilitating the sale of infringing products – not clearly, for example, the platforms the putative 'bot' may be communicating through.[128]

71 However, the provider-user-speaker distinction can collapse in practice. Consider the use of an AI text-generation system such as GPT-3, a tool which extends prompts into elaborate, arbitrary length strings of text,[129] to generate 280 character strings for posting on Twitter. Who is the provider? GPT-3 is an API to a 'raw' model.[130] To comply, should GPT-3 always return a string that ends in "#bot"? Such an interpretation would be far from technology neutral. Should some tendency to convincingly disclose itself as a bot be embedded, through training, in the 175bn parameter model itself? This seems technically daunting for a system that can just as easily produce fake legislation as a fake news article. The Draft AI Act assumes a chatbot vendor pieces together and resells a system, but APIs themselves are becoming user-friendly and intuitively configurable. To make sense, the Draft AI Act could drop its distinction between *user* and *provider*, and think in hybrids, in the style of 'prosumer law' long called for an information regulation.[131]

## 2. Emotion Recognition and Biometric Categorisation Disclosure

72 Users of an emotion recognition or a biometric categorisation system must inform exposed persons of the operation of the system, except in the case of biometric categorisation permitted by law to be used for crime prevention.[132]

73 It is unclear what this provision adds to data protection law. When emotional recognition or biometric categorisation systems process personal data, data protection law requires that users of such systems inform individuals of, *inter alia*, the existence of and purposes of such processing.[133] It is therefore hard to work out what the Commission intended in this provision. Perhaps the Commission intended to mandate clear signage, given users' lack of interest in privacy policies? If so, the Draft AI Act's provision appears ineffective. It is not more strongly worded than the provisions of the GDPR, and the European Data Protection Board already state that users of camera systems must state the purposes on a sign.[134]

74 Perhaps this provision is for where emotional recognition or biometric classification systems do *not* process personal data? Some developers claim this, such as the Fraunhofer Institute's *Anonymous Video Analytics for Retail and Digital Signage* (AVARD) system, which relies on an unpublished assertion to that effect from the Bavarian DPA for the Private Sector.[135] This interpretation brings many problems.[136] Other DPAs, national case law based on the GDPR and scholars claim that personal data is processed in these situations.[137] This reasoning would see the Commission implicitly legitimising a contentious and restrictive reading of the GDPR.

75 Either way, arguing the main issue with emotional or biometric categorisation is a *lack of transparency* risks legitimising a practice with little-to-no scientific basis and potentially unjust societal consequences. A recent literature review concluded that, '[i]t is not possible to confidently infer happiness from a smile, anger from a scowl, or sadness from a frown, as much of current technology tries to do when applying what are mistakenly believed to be the scientific facts'.[138] Those claiming to detect emotion use oversimplified, questionable taxonomies; incorrectly assume universality across cultures and contexts; and

---

124 European Commission, 'Tackling Online Disinformation: A European Approach (COM/2018/236 Final)' (26 April 2018) para 3.1.1; European Commission, 'Code of Practice on Disinformation' (26 September 2018) para 5 https://ec.europa.eu/digital-single-market/en/news/code-practice-disinformation.

125 European Commission, 'Guidance on Strengthening the Code of Practice on Disinformation (COM(2021) 262 Final)' (26 May 2021) 12.

126 European Commission, 'Proposal for a Regulation of the European Parliament and of the Council on a Single Market For Digital Services (Digital Services Act) and Amending Directive 2000/31/EC (COM(2020) 825 Final)' (15 December 2020), recital 69.

127 Madeline Lamo and Ryan Calo, 'Regulating Bot Speech' (2019) 66 UCLA L Rev 988.

128 Regulation (EU) 2019/1020 of the European Parliament and of the Council of 20 June 2019 on market surveillance and compliance of products and amending Directive 2004/42/EC and Regulations (EC) No 765/2008 and (EU) No 305/2011, OJ L 169/1 ('Market Surveillance Regulation'), art 7(2).

129 Tom B Brown and others, 'Language Models Are Few-Shot Learners' [2020] arXiv:200514165 [cs].

130 See generally on AI-as-as-Service, Cobbe and Singh (n 32).

131 See generally Ian Brown and Christopher T Marsden, *Regulating Code: Good Governance and Better Regulation in the Information Age* (MIT Press 2013).

132 AI Act, art 52(b).

133 GDPR, art 13. Under the GDPR, the obligations are imposed on 'data controllers'.

134 European Data Protection Board (n 50) para 116.

135 See Michael Veale, 'Governing Machine Learning that Matters' (PhD, University College London 2019) 217. The report from the Bavarian DPA is on file with the lead author (LDA-1085.4-1368/17-I, dated 8 June 2017). Additional potential examples are given in Damian Clifford, 'The Legal Limits to the Monetisation of Online Emotions' (PhD, KU Leuven 2019) paras 309, 311.

136 Damian George and Kento Reutimann, 'GDPR Bypass by Design? Transient Processing of Data under the GDPR' (2019) 9 International Data Privacy Law 14; Clifford (n 135) paras 309–311.

137 *R (on the application of Edward Bridges) v The Chief Constable of South Wales Police and Secretary of State for the Home Department* [2019] EWHC 2341 (Admin) [59]; Information Commissioner's Office, 'Information Commission's Opinion: The Use of Live Facial Recognition Technology in Public Places' (18 June 2021) 27 https://ico.org.uk/media/for-organisations/documents/2619985/ico-opinion-the-use-of-lfr-in-public-places-20210618.pdf; Peter Alexander Earls Davis, 'Facial Detection and Smart Billboards: Analysing the "Identified" Criterion of Personal Data in the GDPR' (2020) 6 Eur Data Prot L Rev 365.

138 Lisa Feldman Barrett and others, 'Emotional Expressions Reconsidered: Challenges to Inferring Emotion From Human Facial Movements' (2019) 20 Psychological Science in the Public Interest 1, 46.

risk '[taking] us back to the phrenological past' of analysing character traits from facial structures.[139] The Draft AI Act's provisions on emotion recognition and biometric categorisation seem insufficient to mitigate the risks.

### 3. Synthetic Content ('Deep Fake') Disclosure

76 Users of AI systems that generate or manipulate image, audio or video content that appreciably resembles 'existing persons, objects, places or other entities or events' and would falsely appear to a person to be authentic are required to disclose the artificial nature of the resulting content. Exemptions exist for legally authorised crime prevention–related purposes, or necessity to exercise freedom of expression or freedom of the arts and sciences.[140] The narrow definition of 'user' also exempts 'personal non-professional' activities.[141]

77 The mischief this provision tackles is difficult to identify. Convincing likenesses of existing persons may harm important facets of the self,[142] and already trigger some personality protection.[143] Disclosure may only partially assist the subject; the remedy seems to focus instead on protecting the risk of misled audiences. Furthermore, non-human 'entities' do not need persona protection. In most cases, EU law already bans misleading commercial practices likely to be covered by the Act's scope of fake 'persons, objects, places, or other entities or events'.[144] If restrictions against professional deep fakes are required, it may be better to focus enforcement with consumer protection authorities instead of the product safety regulators the Draft AI Act centres upon now.

78 The residual mischief may instead relate to situations where misplaced beliefs of authenticity present danger. For example, AI systems that increase the resolution of images, or generate 3D models from 2D images infer the remainder. People could mistakenly regard such outputs as reliable measurement rather than inference, and such mistakes could cause harm.[145] Yet the Draft AI Act's disclosure obligation falls on the *user*, not the *provider*. If *users* are not aware of synthetic aspects or authenticity-related software limitations, how can they protect the safety of individuals affected by their actions? Where professional users *intentionally* seek to deceive others using software – perhaps producing fake evidence to dispute parking tickets – why not tackle this using conventional laws of fraud?

79 Perhaps this provision seeks to secure some *right to reality* grounded in fundamental rights? This seems reasonable in relation to 'fake news' of salient events, people, places or objects. However, the provision's scope seems too broad. If this provision is attempting to contribute to the regulation of disinformation in media law, it ignores the specificities of each medium. Scholars have noted that legislators do this at their peril.[146] Examples help to highlight the tensions. The provision may also apply to a business using an AI stock image generator to create a bland, original scene of a board room or customer interaction for marketing purposes.[147] Stock photos are rarely of the real businesses in any case, and a generator may end up cheaper, easier and more tailored. Is it reasonable to require disclosure for such synthetic scenes?

80 Finally, as an obligation on *users*, this provision raises the practical enforcement questions comparable to questions regarding bot disclosure laws. How does an enforcement body investigate putatively undisclosed deep fakes? As discussed in section III.4.

c) (para. 58 et seq.) above, it is unclear whether market surveillance authorities have powers to unmask and investigate professional users of platforms who are communicating using an AI system rather than selling one. Moreover, such authorities are unlikely to have the forensics expertise needed for investigating such communications. In sum, the 'deep fake' provision of the Draft AI Act raises many questions.

## V. Harmonisation and Pre-Emption

81 The Draft AI Act aims to 'prevent unilateral Member States actions that risk to fragment [sic] the market and to impose even higher regulatory burdens on operators developing or using AI systems'.[148] Where the Draft AI Act's provisions entail this 'maximum harmonisation', Member States' abilities to act in that area are disabled. Member States must disapply conflicting national rules and accept compliant products on their markets.[149] If a provision is found to not maximally harmonise an area, or only harmonises certain areas, Member States retain competence to adopt more stringent standards. The pre-emptive effect of the Draft AI Act could have far-reaching consequences.

82 Characterising the extent of maximum harmonisation requires identifying the material scope of the instrument (the 'occupied field') and determining the nature of residual Member State

---

competence within it.150 The Draft AI Act lays out 'harmonised rules for the placing on the market, the putting into service and the use of [AI systems] in the Union'.151 The occupied field is thus not Title III 'high risk' systems, but all AI systems. The Draft AI Act defines AI systems by intersecting a functional definition of systems that 'for a given set of human-defined objectives, generate outputs such as content, predictions, recommendations, or decisions influencing the environments they interact with',152 with a descriptive definition based on a wide list of technologies listed in Annex I including 'logic' and 'statistical' approaches. The broad scope might not encompass *all* software, but captures some features of most. All the Draft AI Act's obligations on providers or users relate to significantly narrower subsets of this definition. However, the 'occupied field' with which to examine the pre-emptive effect relates to the broadest definition.

83 The Draft AI Act therefore has an unusual misalignment between the target of its substantive obligations (primarily high-risk systems) and its material scope (all AI systems). Normally, NLF instruments do not adjust requirements (and certainly not regimes) to products of differing risk level, but instead adjust how onerous the conformity assessment 'modules' are (e.g. notified bodies versus internal control).153 NLF instruments do not typically harmonise areas in which they impose no requirements.154 The Draft AI Act, however, seeks to both create harmonised standards, and preclude a broad array of software from further restrictions without imposing any of its own.

84 The way in which the Draft AI Act may restrict further rules on *marketing* and on *use* differ, and so we look at them in turn.

### 1. Marketing

85 Put simply, marketing of all AI systems, not just high-risk systems, is fully harmonised by the Draft AI Act.155 If Member States wish to introduce further restrictions on the placing of *any* AI system on the market, such as to limit carbon footprint or support accessibility,156 they must rely on limited exceptions in Article 114 TFEU, subject to approval by the Commission.157

### 2. Use

86 The extent of the AI Act's pre-emption of national rules on use of AI systems is unclear. NLF instruments before the Draft AI Act have never placed obligations on *users* after installation or first use, so few clear analogies exist.158

### a) Material Scope

87 As a starting point, the material scope of the Draft AI Act is not only certain aspects of use, but concerns 'harmonised rules for [...] the use of artificial intelligence systems',159 and aims to '[prevent] Member States from imposing restrictions on the [...] use of AI systems, unless explicitly authorised by this Regulation.'160 This appears to rule out the possibility that the Draft AI Act is a general 'minimum harmonisation' instrument, setting a horizontal regulatory floor. Such general 'minimum harmonisation' instruments are in any case not permitted by the CJEU under TFEU Article 114.161

88 As an alternate possibility, the Draft AI Act is a *partial harmonisation* instrument, and only transparency rules are harmonised, leaving Member States free to legislate on other issues. The Draft AI Act's scope additionally states it lays down 'harmonised transparency rules for AI systems'. Those rules are listed in Title IV (discussed in section IV. (para. 86 et seq.) above) and concern both use and provision.162

### b) The CJEU Approach

89 In the *Phillip Morris* case, the CJEU took an escape route along those lines. Tobacco firms tried to characterise a Directive as an illegal minimum harmonisation instrument, to limit Member State's residual authority to introduce restrictions such as plain packaging requirements. The Court foiled tobacco firms' attempts by instead interpreting the Directive as only harmonising some areas. However, the Court relied on the Directive's explicit statement in its material scope that it only harmonised 'certain' aspects of packaging and labelling.163 Moreover, the

---

150 Stephen Weatherill, 'The Fundamental Question of Minimum or Maximum Harmonisation' in Sacha Garben and Inge Govaere (eds), *The Internal Market 2.0* (Hart Publishing 2020).
151 AI Act, art 1(1).
152 AI Act, art 3(1).
153 Decision 768/2008/EC, Annex II.
154 The remaining areas are left to TFEU 34 and 36 to govern what remaining restrictions are permitted.
155 With the exception of AI systems developed or used exclusively for military purposes, and, to the criticism of the EDPS and EDPB, to international law enforcement co-operation, which they see as a loophole. See European Data Protection Board and European Data Protection Supervisor (n 47) para 14.
156 See generally Roel Dobbe and Meredith Whittaker, 'AI and Climate Change: How They're Connected, and What We Can Do about It' (*AI Now Institute*, 17 October 2019) https://medium.com/@AINowInstitute/ai-and-climate-change-how-theyre-connected-and-what-we-can-do-about-it-6aa8d0f5b32c accessed 2 July 2021.
157 For example, Member States may still be able to, with the permission of the Commission, introduce a measure relating to 'protection of the environment or the working environment on grounds of a problem specific to that Member State' on the basis of scientific evidence, or 'public health' despite maximum harmonisation, see TFEU, arts 114(5), 114(8).
158 The Commission notes that '[t]he end-user is not one of the economic operators who bear responsibilities under Union harmonisation legislation' in relation to 'any operation or transaction', although this might 'fall under another regulatory regime, in particular at national level.' See Commission Notice of the 27th July 2016 on the 'Blue Guide' on the implementation of EU products rules 2016, OJ C272/1 15. The closest NLF instruments get to placing obligations on users is if incorporating assessment of installation or assembly (e.g. a lift) or distribution (e.g. a measuring instrument) is key to assessing a product, for example a lift or delicate measuring instrument – called 'putting into service'. See Blue Guide, 22.
159 AI Act, art 1(a).
160 AI Act, recital 1.
161 See Case C-547/14 *Philip Morris* ECLI:EU:C:2016:325 [70]–[72] (the Court noting that if it accepted that if a Directive based on TFEU 114 allowed Member States to further legislate in a field *which the Directive had harmonised*, it would have been adopted illegally on that basis). Note however that Weatherill characterises the jurisprudence on minimum harmonisation, particularly regarding other Treaty bases, as 'a mess' and indicates that there is a 'sense that the Court is not fully aware of what it is doing'. See Weatherill (n 150).
162 AI Act, art 1(c).
163 Directive 2014/40/EU of the European Parliament and of the Council of 3 April 2014 on the approximation of the laws, regulations and administrative provisions of the Member States concerning the manufacture, presentation and sale of tobacco and related products and repealing Directive 2001/37/EC, OJ L 127/1 ('Tobacco Products Directive'), art 1(b)

Directive regularly referenced to 'aspects not regulated', and included an explicit clause allowing Member States to go further in some areas.[164]

90 The Draft AI Act lacks all the tools the CJEU relied on to escape total maximum harmonisation. The Act does note that obligations on users of *high-risk systems* are 'without prejudice to other user obligations under Union or national law'.[165] But no similar provision exists applying to the Draft AI Act's entire scope, which itself is broad. There is therefore legal uncertainty whether existing national algorithmic transparency requirements applying beyond 'high-risk' systems, such as public sector provisions in France, may have to be disapplied.[166]

### c) Fragmentation and the Cliff-Edge of Uncertainty

91 Even if a partial, rather than maximum, harmonisation instrument, EU primary law creates opportunities for companies to challenge use restrictions in national law.[167] This is no novelty of the Draft AI Act, and a risk in any non- or partially harmonised area.[168] The Draft AI Act is primarily based on EU free movement competences rather than on fundamental rights. The CJEU finds that use limitations by Member States can constitute a measure equivalent to a quantitative restriction on trade if they directly and substantially impede access to the market.[169] Such measures can be justified on the basis of objective justifications or public interest requirements.[170] However, while many justifications are possible, the Court increasingly requires Member States to justify them in terms of proportionality, fundamental rights and legal certainty.[171] Indeed, in areas where the Draft AI Act is directly relevant, such as AI systems used by employers to manage employees, the CJEU has used freedom of movement law (in what both labour and internal market scholars characterise as troubling misapplications[172]) to strike down collective bargaining efforts, such as in the controversial cases of *Viking* and *Laval*.[173]

92 Some readers may feel the EU should act to prevent fragmented rules disrupting trade of AI systems. This seems defensible for a category of 'high risk' systems for which requirements may be complex. However, the consequence of the Draft AI Act may be to create a stark, arbitrary divide between high-risk systems, which are regulated, and non-high-risk systems, which Member States are effectively forbidden from regulating (or become at-risk of constant challenge). As the Draft AI Act does little to reduce fundamental rights risks of the many systems not covered by Annex III, this 'cliff edge' from some rules to practically none seems difficult to justify.

## VI. Post-Marketing Controls and Enforcement

93 In the last two decades, New Legislative Framework regimes have evolved to include *post*-marketing controls inspired by pharmacovigilance.[174] The Draft AI Act has several components of such regimes.

94 The Draft AI Act gives an important role to *market surveillance authorities* (MSAs). MSAs are public bodies with wide ranging powers to obtain information, apply administrative penalties, withdraw products and oblige intermediaries, including information society services (e.g. online API providers or AI-as-a-Service marketplaces), to cease offering products, or co-operate with MSAs to mitigate their risks.[175]

95 MSAs are typically government departments or regulatory agencies.[176] The Commission does not foresee the 'automatic' creation of any bespoke national authorities,[177] and Member States retain discretion on which authorities will be competent for the new 'standalone' high risk systems in the Draft AI Act. AI systems in scope because they are or are parts of products regulated by harmonised legislation in Annex II are regulated by the MSA for those instruments.[178] In relation to law enforcement users and Union bodies, data protection authorities will gain MSA roles.

96 Penalties are the maximum of 6 % of global turnover or 30m EUR (whichever is higher) for breaches of the Title II prohibitions or Title III data quality requirements; for other rules, the maxima are lower.[179] If the infringer is a public body however, penalties are chosen by Member States.[180]

---

## 1. Notification Obligations and Complaints

97 MSAs' main information source is through a chain of notification obligations. Users of AI systems must monitor it and inform providers of new risks or malfunctions.[181] Providers must tell the MSA if their post-marketing monitoring reveals risks or non-compliance.[182]

### a) Nor Rights for AI-System-Subjects

98 However, individuals affected by AI systems have no right to complain to an MSA in the same way that, for example, they have a right to lodge a complaint to and seek a judicial remedy against a supervisory authority under data protection law.[183] The Draft AI Act creates no legal right to sue a provider or user for failures under the Draft AI Act, although routes may exist for litigants to argue that standards, such as those used in the Draft AI Act, should be considered in national tort cases.[184] The absence of affected individuals and communities in the Draft AI Act is already criticised by the European Data Protection Board and the European Data Protection Supervisor.[185] Collectives such as consumer groups also lack any rights, such as representative complaints possible under the GDPR.[186] In general guidance on MSAs, the European Commission states that Member States '[must] ensure that consumers and other interested parties are given an opportunity to submit complaints [and have them] followed up appropriately'.[187] However, EU law merely requires MSAs to handle complaints competently and to consider complaints like any other information source.[188]

99 Outside the field of NLF rules, *complaint mechanisms* have been pivotal in developing Union case-law where regulators are reticent to challenge the practices of powerful technology firms.[189] As only those with obligations under the Draft AI Act can challenge regulators' decisions, rather than those whose fundamental rights deployed AI systems affect, the Draft AI Act lacks a bottom-up force to hold regulators to account for weak enforcement. Data protection law where affected groups *can* raise complaints is already characterised by inaction and paralysis. Enforcement of the Draft AI Act therefore seems likely to play out in an even more lacklustre way than it has with the GDPR to date.[190]

100 Some (non-NLF) EU product regulation contains complaint handling obligations that the Draft AI Act could learn from. For instance, the EU Timber Regulation ensures that the monitoring authority must accept 'substantiated concerns' from civil society and other groups, and should 'endeavour' to carry out checks on operators when in possession of these.[191]

### b) Incoherence of the Enforcement System

101 To make this worse, the Draft AI Act's enforcement system is set up as NLF enforcement, even though only Title III is an NLF-style regime. The Act's Title II prohibitions and Title IV transparency requirements regulate users, who are brought into scope of MSA powers through a bold interpretative expansion of the MSA Regulation to simply add 'user' to 'economic operator'.[192] Such expansion underestimates how different regulating users will be from normal NLF oversight.

102 Under the Draft AI Act, MSAs are expected, among other obligations, to look for synthetic content on social networks, assess manipulative digital practices of any professional user, and scrutinise the functioning of the digital welfare state. This is *far* from product regulation. MSAs are not guaranteed to be independent of the Government, as a data protection supervisory authority must be.[193] Apart from that, the European Commission estimates the entire enforcement of the Draft AI Act will only take between 1 and 25 extra full-time staff at Member State level.[194] These authors think this is dangerously optimistic.

## 2. Database of Standalone High-Risk AI Systems

103 The Draft AI Act proposes a new, central database, managed by the Commission, for the registration of 'standalone' high-risk AI systems. This approach appears to be modelled after the database and device registration requirements in the new Medical Devices Regulations.[195]

104 The database required by the Draft AI Act makes sense to help MSAs, who otherwise might find locating illicit AI systems difficult. The Commission further proposes to make this database public, to also help 'other people', presumably civil society or journalists, to uncover illicit AI.[196] With the exception of AI systems for law enforcement, migration, asylum and border management, providers must upload electronic instructions for use of AI systems in this database.[197] These instructions state what users must follow to avoid liability under the Draft AI Act. Yet without clear complaint rights, bottom-up enforcement on the basis of this database will be significantly hampered.

---

181 AI Act, art 29(4).
182 AI Act, arts 61(1), 62(1).
183 GDPR, 679, arts 77-78; Law Enforcement Directive, arts 52–53.
184 van Leeuwen (n 103) 20–21.
185 European Data Protection Board and European Data Protection Supervisor (n 47) para 18 (noting 'the absence of any reference in the text to the individual affected by the AI system appears as a blind spot in the Proposal' and the absence of any 'rights and remedies').
186 GDPR, art 80.
187 Blue Guide, 104.
188 Market Surveillance Regulation, arts 11(3)(e), 11(7)(a) (stating MSAs have only loose obligations to establish 'procedures for following up on complaints or reports on issues relating to risks or non-compliance' and to take into account complaints when taking a 'risk-based approach' to decide which checks to undertake).
189 See e.g. Case C-362/14 *Maximillian Schrems v Data Protection Commissioner* ECLI:EU:C:2015:650.
190 See European Parliament resolution of 25 March 2021 on the Commission evaluation report on the implementation of the General Data Protection Regulation two years after its application (2020/2717(RSP)) para 17 (on regulatory paralysis in data protection enforcement).
191 Regulation (EU) No 995/2010 of the European Parliament and of the Council of 20 October 2010 laying down the obligations of operators who place timber and timber products on the market OJ L 295/23, recital 22.
192 AI Act, art 63(1)(a).
193 Charter of Fundamental Rights of the European Union, art 8(3).
194 European Commission, 'AI Act Impact Assessment' (n 95), Annex 3, 25. Note, the larger data protection authorities have hundreds of staff. European Data Protection Board, 'First Overview on the Implementation of the GDPR and the Roles and Means of the National Supervisory Authorities' (Report presented to the European Parliament's Civil Liberties, Justice and Home Affairs Committee (LIBE), 26 February 2019).
195 Regulation (EU) 2017/745 of the European Parliament and of the Council of 5 April 2017 on medical devices, amending Directive 2001/83/EC, Regulation (EC) No 178/2002 and Regulation (EC) No 1223/2009 and repealing Council Directives 90/385/EEC and 93/42/EEC OJ L 117/1 (Medical Devices Regulation), arts 28–29; Regulation (EU) 2017/746 of the European Parliament and of the Council of 5 April 2017 on in vitro diagnostic medical devices and repealing Directive 98/79/EC and Commission Decision 2010/227/EU OJ L 117/176, arts 25–26. This database is called EUDAMED, and designed *inter alia* to track Unique Device Identifiers (UDIs) and bring the EU in line with the US FDA, following requirements in the Food and Drug Administration Amendment Act of 2007. Early data protection laws also had some notification requirements.

105 The database is an interesting innovation for *users* who are also *providers*. If a company internally develops a high-risk AI system (e.g. for hiring) and puts it into service 'for [its] own use',[198] the company is both provider and user. It must declare the system on the database, and upload instructions. This seems like an important tool for accountability, also beyond Draft AI Act requirements, but also something firms may contest in court, claiming violations of trade secrets and the like.

106 Similarly, *users* who disregard the instructions to use an AI system 'off-label', or substantially modify it, also become *providers* and therefore must declare they have done such publicly.[199] However, changes to AI systems which continue to learn within parameters 'pre-determined by the *provider*' do not constitute substantial modification.[200] Many AI systems that are based on machine learning will fall within that exception – but should they? As with the GPT-3 example in section IV.1. (para. 87 et seq.) above, users of general-purpose AI-as-a-Service APIs, designed to be repurposed, changed and configured, may find themselves with conformity assessment obligations without the capacity or expertise to carry them out.

## VII. Concluding Remarks

107 The Draft AI Act is a world-first attempt at horizontal regulation of AI systems. It has many sensible elements, such as differentiating requirements by risk level, introducing prohibitions, and facilitating societal scrutiny via a public database of systems.

108 However, the Draft AI Act also has severe weaknesses. It is stitched together from 1980s product safety regulation, fundamental rights protection, surveillance and consumer protection law. We have sought to illustrate how this patchwork does not make the Draft AI Act comprehensive and watertight. Indeed, these pieces and their interaction may leave the instrument making little sense and impact. The prohibitions range through the fantastical, the legitimising, and the ambiguous. The high-risk regime looks impressive at first glance. But scratching the surface finds arcane electrical standardisation bodies with no fundamental rights experience expected to write the real rules, which providers will quietly self-assess against. The transparency provisions either add little to existing law or raise more questions than answers when their implications are considered. The enforcement mechanism is a creature of product safety. The regime is expected to regulate AI *users* too, yet affected communities are provided with no mechanism for complaint or judicial redress.

109 The pre-emptive effect of the Draft AI Act's maximum harmonisation raises further questions. The Draft AI Act's poor drafting risks an extraordinarily broad scope, with the supremacy of European law restricting legitimate national attempts to manage the social impacts of AI systems' uses in the name of free trade. The Draft AI Act may disapply existing national digital fundamental rights protection. It may prevent future efforts to regulate AI's carbon emissions or apply use restrictions to systems the Draft AI Act does not consider 'high-risk'. Counterintuitively, the Draft AI Act may contribute to deregulation more than it raises the regulatory bar.

110 This paper cannot and has not covered all the Draft AI Act's facets. Many aspects are omitted and deserve further scrutiny. We urge legislators and civil society to rise to this challenge, and hope that we have demonstrated some of the complexities making this a particularly important instrument to analyse throughout its legislative process.

*Dr. Michael Veale*

Lecturer in Digital Rights and Regulation, Faculty of Laws, University College London

Data protection law, human-computer interaction, machine learning

m.veale@ucl.ac.uk

https://michae.lv

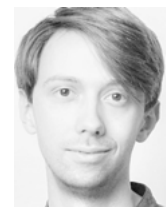

*Prof. Dr. Frederik Zuiderveen Borgesius, LL.M.*

Interdisciplinary Hub for Security, Privacy and Data Governance, Radboud University, The Netherlands

Privacy, Discrimination and generally human rights in the context of new technologies

frederikzb@cs.ru.nl

https://www.ru.nl/english/people/zuiderveen-borgesius-f

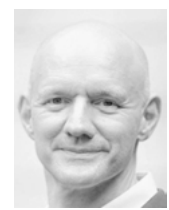

---

196 European Commission, 'AI Act Impact Assessment' (n 95) 56. Databases mandated by law in fields such as trademarks have been discussed as potentially helpful to algorithmic accountability. See e.g. Amanda Levendowski, 'Trademarks as Surveillance Transparency' (2021) 36 Berkeley Tech L J__.
197 AI Act, Annex VIII, para 11.
198 AI Act, art 3(11).
199 AI Act, art 28.
200 AI Act, art 43(4).